\documentstyle[11pt,newpasp,twoside,psfig]{article}
\index{summary}
\index{instructions}
\index{template}

\def\edcomment#1{\iffalse\marginpar{\raggedright\sl#1\/}\else\relax\fi}
\reversemarginpar

\begin{document}
\title{High time resolution observations of a Vela glitch}
\author{R Dodson, D.R. Lewis, P.M. McCulloch}
\affil{School of Maths \& Physics, University of Tasmania, Australia}

\begin{abstract}
  Pulsars are rotating neutron stars, sweeping the emission regions
  from the magnetic poles across our line of sight. Isolated neutron
  stars lose angular momentum through dipole radiation and (possibly)
  particle winds, hence they slow down extremely steadily, making them
  amongst the most reliable timing sources available. However, it is
  well known that younger pulsars can suffer glitches, when they
  suddenly deviate from their stable rotation period. On 2000 January
  16 (MJD 51559) the rate of pulsation from the Vela pulsar (B0833-45)
  showed such a fractional period change of {\rm $3.1\times 10^{-6}$},
  the largest recorded for this pulsar. The glitch was detected and
  reported by the Hobart radio telescope. The speedy announcement
  allowed the X-ray telescope, Chandra, and others, to make Target of
  Opportunity observations. The data placed an upper limit of 40
  seconds for the transition time from the orginal to the new
  period. Four relaxation timescales are found, which are believed to
  be due to the transfer of inertia through the internal
  structure. One is very short, about 60 seconds; the others have been
  previously reported and are 0.56, 3.33 and 19.1 days in length.

\end{abstract}

\section{Introduction}

Observations of pulsar glitches, in addition to providing insights
into the phenomenon itself, offer one of the few probes of neutron
star structure, and thus the physics of ultra-dense matter. Vela is
the brightest known radio pulsar, and as it is at a declination of
-45$^o$, it is above the horizon at the Hobart Radio Observatory
(Mount Pleasant) for more than 18 hours a day.  It undergoes large
glitches in pulse rate every few years and so provides an excellent
test-bed for the neutron star equations of state.

Since 1981 the University of Tasmania has devoted a 14m diameter
antenna at its Mt Pleasant Observatory to measurements of arrival
times of pulses from the Vela pulsar. During this time we have
observed 7 large glitches, or sudden decreases, in the period of the
pulsar (McCulloch et al 1983, McCulloch et al 1987, McCulloch et al
1990, McCulloch 1996)

The telescope has a single pulse observing system whose speed and
sensitivity have been enhanced in order to answer a number of
questions; how quickly does the crust accelerate to the new period
during a glitch, how soon does the recovery from the glitch start,
and what is the form of this recovery?

On 2000 January 16 (MJD 51559) the rate of pulsation jumped with a
fractional period change of {\rm $3.1\times 10^{-6}$}, the largest
recorded for this pulsar. The glitch was automatically detected and we
issued an IAU telegram (Dodson, McCulloch, \& Costa 2000) within 12
hours, allowing the X-ray telescope, Chandra to make Target of
Opportunity observations (Helfand, Gotthelf, \& Halpern 2001) . These
observations have so far failed to find the signature of neutron star
heating, which was the driver for the observations, but have produced
spectacular images of the X-ray pulsar wind nebula.

\section{Observations}

The reported observations were made simultaneously at 635~MHz, 990~MHz
and 1390~MHz, to allow continuous measurement of the dispersion
measure (DM). The signal-to-noise ratio in each total intensity
profile is typically 30:1, allowing a mean pulse arrival time to be
determined to an accuracy of $80\mu$s at 635~MHz, $60\mu$s at 990~MHz
and $180\mu$s at 1390~MHz per integration.

The recent improvements in the time resolution have been achieved by
de-dispersing over 8 adjacent channels at 990~MHz, thereby increasing
the signal to noise ratio and allowing observations of single
pulses. The de-dispersed bandpass is sampled at 2 kHz and recorded
directly onto disk for later retrieval, while an `on the fly'
monitoring system folds on a ten second basis for which the RMS 
is $85\mu$s. These arrival times are monitored and if a glitch is
detected a warning is issued and the single pulse data are retained.

\section{Timing fits}

The data presented here were recorded between MJD 51505 and 51650.
The arrival times have been transformed to the Solar
System barycentre using standard techniques.  The position and proper
motion of the Vela pulsar was defined by data from  the Radio VLBI
position of Legge (Legge 2001). 

The recorded TOAs from all frequencies and both systems were fitted in
the program {\bf TEMPO} (Taylor et al 1970).  The results of this fit are
given in Table 1.

Shortly after 07:34 UT, the residuals diverge from the fit, indicating
a sudden decrease in pulse period. The period jump occurs
on a very short timescale, without warning. The observations are
consistent with an instantaneous change in period; modelling has shown
that a spin-up timescale of forty seconds would produce a three sigma
signal.

The separation into four time-scales is clear. The longer three decay
terms are similar to those previously reported (Alpar et al 1993,
Flanagan 1990), and are in an approximately equal ratio of
5.9:5.7. These have been associated with the vortex creep models by
(Alpar et al 1993) and others. The fast decay timescale, not
previously observed (or observable) is shown separated from the other
effects in figure 1. We have subtracted the terms found by {\bf TEMPO}
in the 2 minute data from the single pulse data folded for 10
seconds. In this plot a gradual spin-up (as opposed to an
instantaneous) would be a negitive excursion around the projected time
of the glitch, as we'd have overestimated the phase in the model. We
see a positive excursion, indicating that the true glitch epoch was
later, and is followed by a rapid decay. We have fitted a linear rise
($\Delta \nu \Delta t$) followed by a forth decay term to this.

\section{Future development}

Since the acceleration of the crust cannot be instantaneous, it should
be possible to observe the spin-up of the rotation period of the
pulsar. The parallel single pulse system designed to observe this has
a three $\sigma$ detection limit of less than 40 seconds. This is in
contrast to the observations made on the Crab (Wong, Backer, \& Lyne
2001) where the spin-up timescale has been observed to take about
$\sim 0.5$ of a day. Further improvements of the single pulse system
are being undertaken which will allow a detection limit of a few
seconds, which is of the order of the fastest coupling times in all
EOS models.  Higher time resolution will allow further constraints on
the coupling of the crust to the liquid interior, including the core.

\vspace{2cm}
\noindent

\begin{table}[h]
\sffamily
\begin{center}
\begin{tabular}{|c|c|c|}
\hline
\multicolumn{3}{|c|}{Parameters for Epoch 51559}\\
\hline
$\nu/Hz$ & $\dot{\nu}/Hz~s^{-1}$ & $\ddot{\nu}/Hz~s^{-2}$ \\
\hline
11.194615396005 & -1.55615E-11 & 1.028E-21 \\
\hline
\end{tabular}

\begin{tabular}{|l|l|}
$\Delta \nu_p/Hz$ & $\Delta \dot\nu_p/Hz~s^{-1}$ \\
\hline
3.45435(5)E-05&-1.0482(2)E-13\\
\hline
$\tau_n$ & $\Delta \nu_n/10^{-6} Hz$\\
\hline
$1.2\pm0.2$mins & 0.020(5)\\ 
00.53(3) days& 0.31(2)\\ 
03.29(3) days& 0.193(2)\\ 
19.07(2) days& 0.2362(2)\\ 
\hline
DM & 67.99 \\
\hline
\end{tabular}
\end{center}
\caption{Parameters for the glitch epoch 51559.3190. Errors are the
one sigma values.  The data fit is from MJD 51505 through to 51650
(November 1999 to April 2000).}

\label{tab:parameters}
\end{table}

\begin{center}
\begin{figure}
\psfig{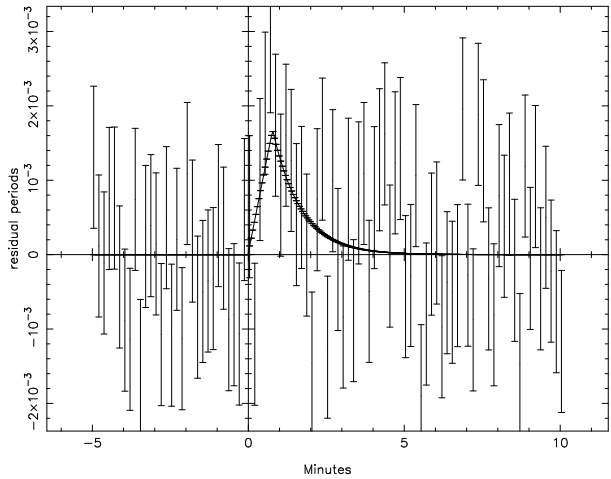}
\caption{The previously unobserved fast decay, 
with ten second folds. The other decay terms are removed revealing the later 
start, and decay of the fastest term.} 
\end{figure}
\end{center}

\end{document}